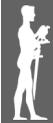

# CAUSES OF FAILURE OF THE PHILLIPS CURVE: DOES TRANQUILLITY OF ECONOMIC ENVIRONMENT MATTER? [1]

**Yhlas Sovbetov, Muhittin Kaplan**

Istanbul University, Turkey

**Abstract:**

Although empirical literature regarding the Phillips curve is sizeable enough, there is still no wide consensus on its validity and stability. The literature shows that the Phillips relationship is fragile and varies across countries and time periods; a statistical relationship that appears strong during one decade (country) may be weak the next (other). This variability might have some grounds for idiosyncrasy of a country and its economic environment. To address it, this paper scrutinizes the Phillips relationship over 41 countries over the period 1980-2016, paying attention to how inflation dynamics behave during tranquil and recessionary periods. As a result, the paper confirms the variability of the Phillips relationship across countries, as well as time periods. It documents that the relationship holds in the majority of developed countries, while it fails to hold in emerging and frontier economies during tranquil periods. On the other hand, the relationship totally collapses during recessionary periods, even in developed markets. This shows that tranquillity of economic environment is significantly important for the Phillip trade-off to work smoothly. Moreover, both backward- and forward-looking fractions of inflation remarkably increase during recessionary periods as a result of the Phillips coefficient loses its significance within the model. This indicates that markets become more inflation-sensitive during these periods.

*E-mail: yhlas.sovbetov@ogr.iu.edu.tr

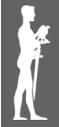

## INTRODUCTION

The empirical study of Alban William Phillips (1958) on change in wage of inflation and fluctuations in unemployment has greatly influenced the course of macroeconomics. He discovered a strong negative trade-off relationship between unemployment and inflation in the UK over the 1861-1957 period, which is today known as the "Phillips Curve". Despite some early criticisms of the basic tenets of the *Phillips Curve* (Samuelson and Solow, 1960; Phelps, 1967; Friedman, 1968; Lucas, 1976), the hypothesis remains one of the most important foundations for macroeconomics. However, after 2008 Great Recession, many studies have challenged the validity of the Phillips curve (Ball and Mazumder, 2011; Russel and Banerjee, 2008; Paul, 2009; Bernanke, 2010; Karan Singh et al, 2011; Hall, 2011; Daly et al., 2012; Ojapinwa and Esan, 2013; Nub, 2013; Wimanda et al, 2013; Simionescu, 2014; Coibion and Gorodnichenko, 2015; Friedrich, 2016; IMF, 2013; Doser et al., 2017, Sovbetov and Kaplan, 2019), when the unemployment rate rapidly scaled up, but inflation did not decline as much as the curve predicted it would[2] . They also underline the variability of the Phillips relationship across countries.

Russel and Banerjee (2008), Paul (2009), Fendel et al. (2011), Ojapinwa and Esan (2013), Nub (2013), Simionescu (2014), Coibion and Gorodnichenko (2015), and Murphy (2018) investigate nonlinearities in the Phillips Curve caused by external factors. This is quite important as any significant changes in behaviour of inflation and unemployment during recessionary periods might also be one of the reasons for the failure of the relationship. However, the mentioned studies do not focus on behaviour of the Phillips Curve during recessionary periods. To our best knowledge, the extant literature lacks empirical research that examines behavioural changes of the Phillips relationship during recessionary and tranquil periods.

To fill this gap, this research examines the Phillips curve with an up-to-date data over the 1980-2016 period, focusing on tranquil and recessionary periods separately. In order to addresses the variability of the Phillips relationship, this examination has been carried out across 41 different countries from developed, emerging, and frontier markets. For robustness, the research also considers both backward- and forward-looking versions of expectation-augmented Phillips model (EAPC). The study mainly pursues answers for two questions: (1) is the Phillips relationship still valid? (2) Is there any significant change in the Phillips relationship during recessionary and tranquil periods?

The rest of the paper is structured in the following order: the second section briefly reviews the formulation of the Phillips curve and covers the main causes behind its failure during the Great Recession in 2008. The third section describes the data and specifies the methodology for this study. The fourth section presents the findings and interprets them thoroughly, while the final section provides conclusions.

## LITERATURE REVIEW

The Great Recession in 2008 has rekindled interest in the Phillips curve with a particular focus on causes of failure of the empirical relationship between inflation and unemployment, and on "*missing disinflation*". Bernanke (2010) argues that the main causes of the absence of disinflation are well-anchored expectations of households, which were imposed by highly credible Central banks for a long time. His "*anchored expectations*" hypothesis basically states that the credibility of modern central banks has convinced people for a long-run stability in inflation. However, this hypothesis would work only for countries where these two conditions are valid: (1) The country's central bank is highly credible; (2) the impact of external shocks is ineffective to households' budgets. If one of these two conditions fails, then the "*anchored expectations*" hypothesis might generate more significant disinflations.

---

2    This phenomenon is often referred to as "*missing disinflation*".

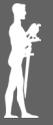

Ball and Mazumder (2011, 2015) also support Bernanke's hypothesis. They attempt to explain the failure of the Phillips Curve in the U.S. throughout 2007-2010 by assuming that inflation expectations are fully anchored at the Federal Reserve's target, and that the level of short-term unemployment captures the labour-market slack. However, Coibion and Gorodnichenko (2015) argue that the "*anchored expectations*" hypothesis was not case, as oil prices were very undulant during 2007-2009 periods. According to West Texas Intermediate (WTI or NYMEX) data, the crude oil prices per barrel was 72 USD in January 2007, when it skyrocketed to historically high record levels of 161 USD within just a year, in June 2008. During the following six months, it follows a decreasing trend until the prices drop to minimum level of 49 USD in January 2009. Afterwards, it once again peaks in June 2009, at 81 USD level. Coibion and Gorodnichenko (2015) argue that the households' expectations are scaled up during the Great Recession in response to sharp increases in oil prices. The households are more responsive to oil prices when they form their future inflation expectations when comparing to professional analysts. The survey-based measures of their expectations reflect changes in the price of their own consumption bundles rather than the overall inflation in the economy. Since gasoline remains as a large portion of the consumption of their income, they adjust their position according to the oil price changes. Therefore, the increased household expectations hindered the downward adjustment of prices, and caused divergence between future inflation expectations of the households and professional analysts. The authors believe it was the key reason behind the absence of disinflation during this period, and they advise analysts to use a better proxy for expected inflation, referring to the Michigan Survey of Consumers (MSC) dataset.

Doser et al (2017) also find that consumer expectations of inflation played an important role during the recent missing disinflation, however, they urge that nonlinearities in the Phillips curve are another reason behind its failure during the Great Recession. They argue that the higher unemployment that emerged due to recessions might not lead to a sharp decrease in inflation. They add that the original 1970's Phillips curve was a convex curve, not a linear relationship. Thus, when unemployment is already high, a further increase in unemployment leads to a smaller disinflation when compared to the case when unemployment was at its historical average.

On the other hand, Russell and Banerjee (2008) study the NAIRU Phillips curve under non-stationarity conditions in the series. They argue that the non-stationarity of inflation having a time-varying mean might be one of the key reasons behind the failure of the Phillips curve. Moreover, a positive relationship might also emerge between inflation and the unemployment rate in the long-run. Ojapinwa and Esan (2013) and Simionescu (2014) further observe a weak positive relationship between inflation and unemployment in Nigeria and Romania, respectively. These findings clearly show that the Phillips trade-off might not always behave as in theory.

Del Boca et al (2010) study the Phillips relationship in Italy and find that the trade-off breaks down during periods of high inflation and unstable macroeconomic environment. They believe that the comparative disadvantage of the Italian economy to withstand adverse supply-side shocks and the poor quality of monetary policy are key reasons of this failure. Nub (2013) explores the Phillips trade-off in Germany with an updated data over the period from 1970 to 2012, and fails to detect a significant negative short-run trade-off. He argues that it happens due to influences of some other factors on the behaviour of inflation and employment. For instance, European Monetary Union (EMU) policies might be one of the reasons behind these influences. Although the EMU eliminates inflation bias due to countries' policy credibility problems (Clerc, Dellas, and Loisel, 2010), it might cause a strong form of nominal rigidities. Nub (2013) believes that negative spillovers coming from other members of the EU might also cause a shock to Germany through the EMU.

Paul (2009) documents the failure of the Phillips trade-off in India. He addresses the liberalization-policy of the early 1990s and supply shocks, such as droughts and oil prices as the main reasons behind this failure. Although he argues that the trade-off might work in the short-run once these shocks are taken into consideration in the model, he believes that this relationship is often evasive or absent in less-developed economies. Sovbetov and Kaplan (2019) also point out that in less-developed and crisis-prone markets the Phillips curve might not work smoothly due to a lack of well-established and freely operating structure of macroeconomic foundations and motivations, as well as a lack of economic environment tranquillity.

In respect of the above-mentioned studies, one can infer that shocks in expectations might cause failure of the Phillips relationship. And sharp changes in expectations tend to occur during an unstable economic environment (recessionary or post-recessionary recovery periods). Therefore, the tranquillity of economic environment should matter in terms of stability and validity of the Phillips relationship. In this regard, this research aims to contribute to the aforementioned field of the Phillips curve literature by examining the relationship during recessionary and tranquil periods separately over 41 different countries from developed, emerging, and frontier markets.

## DATA AND METHODOLOGY

Following Sovbetov and Kaplan (2019), the base equation of this research is formed as below. This odel is specified in order to empirically examine the validity of the backward- and forward-looking Neo-Classical Phillips Curve (NCPC) over Q1:1980-Q1:2016 across 41 countries (Appendix-A) during tranquil and recessionary economic periods.

$$\pi_t = \beta_0 + \beta_1 \pi_t^e + \beta_2 (u_t - u^*) + \beta_3 \pi_t^e DUMMY_\mu + \beta_4 (u_t - u^*) DUMMY_\mu + \varepsilon_t \quad (1)$$

where *DUMMY* is a dummy variable that gets a value of 1 during recessionary periods and zero during other periods; $\pi_t$ and $\pi_t^e$ are proxied by the first differences of the logarithm *CPI* and expected *CPI inflation* over one year respectively; $U_t$, and $U^*$ are proxied by unemployment rate and *NAIRU* respectively in logarithmic form. The $\beta_2$ and $\beta_2+\beta_4$ indicate the Phillips coefficients during tranquil and recessionary periods, respectively. Similarly, $\beta_1$ and $\beta_1+\beta_3$ show fractions of $\pi^e$ in the current inflation during tranquil and recessionary periods respectively. Note that if $\pi_t^e$ equates to $\pi_{t-1}$ then the model converts to NCPC with backward-looking specification. And if it equates to the expected inflation *Et(πt+1)*, then the model takes the shape of NCPC with forward-looking specification. In addition, the study uses manual calculations of standard errors $\beta_2+\beta_4$ and $\beta_1+\beta_3$ with formulas of

$$SE_{\beta2-\beta4} = \sqrt{SE_{\beta2}^2 + SE_{\beta4}^2} \quad \text{and} \quad SE_{\beta1-\beta3} = \sqrt{SE_{\beta1}^2 + SE_{\beta3}^2} \quad {}^3 \text{in order to find out the}$$

significance of the Phillips coefficients during recessionary periods.

Data for these variables are obtained from Thomson Reuters Eikon Datastream, and a fixed constant term is added to all series to handle negative values during the transformation into the logarithmic form, which only shifts $\beta_0$ up, leaving other variables unaffected.

---

3    Cov(β2+β4) and Cov(β1+β3) is approximately zero.

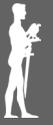

The expected inflation data also retrieved from the Thomson Reuters Eikon Datastream, where the data is formed by their own forecasting survey methodology. The complete raw data used in this study is made available publicly via the following link: https://data.mendeley.com/datasets/8v2mpt7dtp/1, making the results of this study reproducible.

The definitions of recessionary and tranquil periods is quite important in this study. With very basic approach, this study refers all non-growing economic periods as "*recessionary periods*," while remaining (growing economic periods) as "*tranquil periods*." Because various aspects of the economy are disrupted during economic recessions, the study aims to capture all their influences over the Phillips relationship through changed expectations. Alternatively, the study could also just focus on currency crashes that might also fairly proxy the shocks during inflation dynamics. It might, however, be an insufficient approach in some cases in which shocks in inflation fail to visibly influence expectations.

Table 1 provides an overview for the main features of the dataset. The first and second columns of the table present current and expected inflation of CPI derived from Thomson Eikon Datastream. The third column is derived by HP filtering [4] the current inflation (first column) with lambda 1600. Comparing Thomson's forecasted expected inflation, the HP filter derives more accurate estimation. Thus, the table presents NAIRU figures in the fifth column that are obtained using HP-filtering methodology.

**Table 1.** *Overview of Mean and Standard Deviation of Variables*

|  | Current Inflation | | Expected Inflation | | HP InflationTrend | | UnemploymetRate | | NAIRU | | UnemploymentGap | |
|---|---|---|---|---|---|---|---|---|---|---|---|---|
|  | Mean | StdDev. | Mean | StdDev. | Mean | StdDev. | Mean | StdDev. | Mean | StdDev. | Mean | StdDev. |
| AG | 0.0115 | 0.0151 | 0.0197 | 0.0253 | 0.0176 | 0.0283 | 0.1175 | 0.0447 | 0.1169 | 0.0378 | 0.0006 | 0.0191 |
| AU | 0.0027 | 0.0023 | 0.0031 | 0.0008 | 0.0028 | 0.0005 | 0.0645 | 0.0171 | 0.0644 | 0.0160 | 0.0001 | 0.0043 |
| BD | 0.0019 | 0.0018 | 0.0022 | 0.0010 | 0.0019 | 0.0009 | 0.0910 | 0.0167 | 0.0905 | 0.0151 | 0.0005 | 0.0056 |
| BG | 0.0021 | 0.0020 | 0.0022 | 0.0009 | 0.0021 | 0.0005 | 0.0824 | 0.0089 | 0.0817 | 0.0052 | 0.0007 | 0.0054 |
| BR | 0.0476 | 0.1053 | 0.0867 | 0.1978 | 0.0464 | 0.0808 | 0.1145 | 0.0298 | 0.1147 | 0.0267 | -0.0002 | 0.0106 |
| CH | 0.0045 | 0.0072 | 0.0058 | 0.0055 | 0.0044 | 0.0042 | 0.0368 | 0.0059 | 0.0368 | 0.0057 | 0.0000 | 0.0011 |
| CL | 0.0056 | 0.0052 | 0.0061 | 0.0043 | 0.0057 | 0.0040 | 0.0866 | 0.0196 | 0.0865 | 0.0157 | 0.0001 | 0.0100 |
| CN | 0.0019 | 0.0018 | 0.0023 | 0.0008 | 0.0020 | 0.0005 | 0.0780 | 0.0136 | 0.0779 | 0.0116 | 0.0001 | 0.0051 |
| CZ | 0.0054 | 0.0076 | 0.0068 | 0.0076 | 0.0054 | 0.0045 | 0.0545 | 0.0187 | 0.0545 | 0.0168 | 0.0000 | 0.0061 |
| DK | 0.0020 | 0.0019 | 0.0024 | 0.0007 | 0.0020 | 0.0006 | 0.0662 | 0.0261 | 0.0657 | 0.0224 | 0.0005 | 0.0076 |
| ES | 0.0029 | 0.0024 | 0.0032 | 0.0016 | 0.0029 | 0.0015 | 0.1754 | 0.0588 | 0.1736 | 0.0523 | 0.0017 | 0.0142 |
| FN | 0.0017 | 0.0020 | 0.0022 | 0.0010 | 0.0018 | 0.0007 | 0.1000 | 0.0294 | 0.0983 | 0.0247 | 0.0016 | 0.0095 |
| FR | 0.0016 | 0.0017 | 0.0019 | 0.0007 | 0.0016 | 0.0007 | 0.0899 | 0.0100 | 0.0895 | 0.0078 | 0.0004 | 0.0047 |
| GR | 0.0048 | 0.0079 | 0.0053 | 0.0049 | 0.0048 | 0.0046 | 0.1466 | 0.0603 | 0.1458 | 0.0556 | 0.0007 | 0.0155 |
| HN | 0.0108 | 0.0116 | 0.0114 | 0.0092 | 0.0110 | 0.0075 | 0.0773 | 0.0179 | 0.0773 | 0.0152 | 0.0001 | 0.0052 |
| ID | 0.0101 | 0.0134 | 0.0109 | 0.0121 | 0.0101 | 0.0042 | 0.0668 | 0.0191 | 0.0667 | 0.0174 | 0.0001 | 0.0065 |
| IN | 0.0077 | 0.0069 | 0.0078 | 0.0025 | 0.0077 | 0.0021 | 0.0824 | 0.0210 | 0.0823 | 0.0209 | 0.0000 | 0.0005 |
| IR | 0.0022 | 0.0029 | 0.0027 | 0.0016 | 0.0022 | 0.0014 | 0.0902 | 0.0417 | 0.0897 | 0.0366 | 0.0005 | 0.0105 |
| IT | 0.0026 | 0.0019 | 0.0030 | 0.0016 | 0.0027 | 0.0014 | 0.0957 | 0.0188 | 0.0954 | 0.0170 | 0.0003 | 0.0050 |
| JP | 0.0003 | 0.0025 | 0.0006 | 0.0010 | 0.0003 | 0.0007 | 0.0414 | 0.0080 | 0.0413 | 0.0070 | 0.0001 | 0.0028 |

---

4   HP is a technique used to derive long-run levels of variables. The λ is a smoothing parameter that is set by using the Ravn and Uhliq (2002) frequency rule: the number of periods per year divided by 4, raised to the power of x, and multiplied by 1600. Hodrick and Prescott (1997) recommend the value 2 for x, whereas Ravn and Uhliq (2002) suggest using 4 for x. Following Hodrick and Prescott (1997), we derive λ=1600 for our dataset.

| KO | 0.0037 | 0.0034 | 0.0049 | 0.0022 | 0.0038 | 0.0016 | 0.0359 | 0.0121 | 0.0359 | 0.0062 | 0.0000 | 0.0087 |
| MX | 0.0095 | 0.0103 | 0.0105 | 0.0096 | 0.0096 | 0.0067 | 0.0395 | 0.0106 | 0.0392 | 0.0076 | 0.0003 | 0.0061 |
| MY | 0.0030 | 0.0029 | 0.0042 | 0.0015 | 0.0029 | 0.0007 | 0.0325 | 0.0043 | 0.0324 | 0.0022 | 0.0001 | 0.0033 |
| NL | 0.0022 | 0.0021 | 0.0024 | 0.0008 | 0.0022 | 0.0006 | 0.0627 | 0.0165 | 0.0620 | 0.0123 | 0.0007 | 0.0074 |
| NW | 0.0022 | 0.0022 | 0.0024 | 0.0007 | 0.0022 | 0.0003 | 0.0388 | 0.0088 | 0.0389 | 0.0068 | 0.0000 | 0.0039 |
| OE | 0.0021 | 0.0026 | 0.0023 | 0.0008 | 0.0021 | 0.0007 | 0.0475 | 0.0062 | 0.0475 | 0.0045 | 0.0001 | 0.0036 |
| PH | 0.0058 | 0.0043 | 0.0069 | 0.0030 | 0.0060 | 0.0023 | 0.0889 | 0.0184 | 0.0890 | 0.0161 | -0.0001 | 0.0068 |
| PO | 0.0088 | 0.0112 | 0.0118 | 0.0147 | 0.0091 | 0.0111 | 0.1410 | 0.0311 | 0.1406 | 0.0216 | 0.0004 | 0.0145 |
| PT | 0.0032 | 0.0028 | 0.0036 | 0.0027 | 0.0032 | 0.0022 | 0.0831 | 0.0359 | 0.0825 | 0.0337 | 0.0007 | 0.0095 |
| RM | 0.0337 | 0.0456 | 0.0343 | 0.0449 | 0.0300 | 0.0297 | 0.0736 | 0.0243 | 0.0725 | 0.0193 | 0.0011 | 0.0131 |
| RS | 0.0481 | 0.0941 | 0.0761 | 0.1456 | 0.0449 | 0.0577 | 0.0777 | 0.0219 | 0.0777 | 0.0187 | 0.0001 | 0.0080 |
| SA | 0.0072 | 0.0043 | 0.0081 | 0.0029 | 0.0072 | 0.0021 | 0.2334 | 0.0324 | 0.2337 | 0.0278 | -0.0004 | 0.0109 |
| SD | 0.0015 | 0.0028 | 0.0022 | 0.0015 | 0.0016 | 0.0014 | 0.0825 | 0.0191 | 0.0812 | 0.0145 | 0.0013 | 0.0085 |
| SP | 0.0019 | 0.0028 | 0.0026 | 0.0012 | 0.0019 | 0.0013 | 0.0238 | 0.0074 | 0.0236 | 0.0059 | 0.0002 | 0.0043 |
| SW | 0.0009 | 0.0026 | 0.0014 | 0.0013 | 0.0010 | 0.0012 | 0.0342 | 0.0090 | 0.0335 | 0.0049 | 0.0006 | 0.0056 |
| TH | 0.0033 | 0.0042 | 0.0044 | 0.0023 | 0.0033 | 0.0016 | 0.0181 | 0.0113 | 0.0181 | 0.0085 | 0.0000 | 0.0063 |
| TK | 0.0321 | 0.0292 | 0.0380 | 0.0319 | 0.0322 | 0.0251 | 0.0850 | 0.0190 | 0.0848 | 0.0158 | 0.0002 | 0.0091 |
| TW | 0.0016 | 0.0041 | 0.0024 | 0.0014 | 0.0016 | 0.0013 | 0.0372 | 0.0119 | 0.0373 | 0.0103 | 0.0000 | 0.0047 |
| UK | 0.0024 | 0.0029 | 0.0031 | 0.0019 | 0.0024 | 0.0012 | 0.0440 | 0.0204 | 0.0435 | 0.0177 | 0.0005 | 0.0049 |
| US | 0.0025 | 0.0021 | 0.0028 | 0.0007 | 0.0025 | 0.0007 | 0.0601 | 0.0161 | 0.0602 | 0.0122 | -0.0001 | 0.0072 |
| VE | 0.0354 | 0.0265 | 0.0408 | 0.0279 | 0.0358 | 0.0199 | 0.1059 | 0.0335 | 0.1063 | 0.0278 | -0.0004 | 0.0143 |

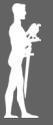

## RESULTS AND DISCUSSION

The estimation of the results of the Eq.1 model and the manual computation of coefficients and standard errors of Phillips coefficients during recessionary periods are displayed at Table 1, where superscripts "N" and "R" indicate estimations for tranquil (normal) and recessionary periods, respectively.

**Table 1.** *Backward/Forward-Looking NCPC Model during Tranquil/ Recessionary Periods*

| Market Class. | Countries | Backward-Looking PC Model | | | | | Forward-Looking PC Model | | | | | Obs |
|---|---|---|---|---|---|---|---|---|---|---|---|---|
| | | $^N\pi_{t-1}$ | $^N U_{GAP}$ | $^R\pi_{t-1}$ | $^R U_{GAP}$ | C | $^N E_t(\pi_{t+1})$ | $^N U_{GAP}$ | $^R E_t(\pi_{t+1})$ | $^R U_{GAP}$ | C | |
| D | Australia | 0.5645*** (0.0739) | -0.0874* (0.0486) | 0.4435*** (0.1757) | -0.1620 (0.1819) | 0.0017*** (0.0004) | 0.6755*** (0.1147) | -0.0900* (0.0476) | 0.7845*** (0.2233) | -0.0344 (0.1456) | 0.0015*** (0.0006) | 145 |
| D | Austria | 0.4559*** (0.0942) | -0.0829* (0.0427) | 0.4582*** (0.1388) | -0.3037** (0.1341) | 0.0012*** (0.0002) | 0.5608*** (0.0845) | -0.0607** (0.0285) | 0.6664*** (0.1323) | -0.2564*** (0.0982) | 0.0009*** (0.0002) | 145 |
| D | Belgium | 0.5603*** (0.0927) | 0.0184 (0.0363) | 0.5982*** (0.1790) | 0.1200 (0.0793) | 0.0011*** (0.0002) | 0.7012*** (0.0808) | -0.0363* (0.0197) | 0.7783*** (0.0926) | -0.0134 (0.0697) | 0.0005** (0.0002) | 145 |
| D | Canada | 0.5248*** (0.1034) | -0.0883** (0.0389) | 0.5856*** (0.1058) | -0.1611 (0.1082) | 0.0013*** (0.0003) | 0.7120*** (0.0660) | -0.0672** (0.0328) | 0.8435*** (0.1354) | -0.0490 (0.1521) | 0.0002 (0.0003) | 145 |
| D | Denmark | 0.5514*** (0.0811) | -0.0342 (0.0247) | 0.6855*** (0.1688) | 0.0196 (0.0472) | 0.0014*** (0.0003) | 0.5598*** (0.1166) | -0.0219* (0.0113) | 0.7644*** (0.1783) | -0.0324 (0.0416) | 0.0004 (0.0004) | 145 |
| D | Finland | 0.6161*** (0.0678) | -0.0154 (0.0341) | 0.7213*** (0.1417) | -0.0588 (0.0574) | 0.0010*** (0.0003) | 0.5473*** (0.0620) | -0.0641*** (0.0207) | 0.7787*** (0.1770) | -0.0648 (0.0427) | 0.0003 (0.0003) | 145 |
| D | France | 0.5113*** (0.0488) | -0.0614* (0.0366) | 0.5886*** (0.0913) | 0.0760 (0.1682) | 0.0003* (0.0001) | 0.5955*** (0.0355) | -0.0444** (0.0214) | 0.8046*** (0.1194) | 0.0331 (0.2915) | -0.0001 (0.0002) | 145 |
| D | Germany | 0.4294** (0.1832) | -0.0921* (0.0516) | 0.4958* (0.2037) | -0.1189 (0.0768) | 0.0026*** (0.0006) | 0.5853*** (0.1047) | -0.0341** (0.0178) | 0.7944*** (0.1535) | -0.0308 (0.0409) | 0.0006* (0.0004) | 145 |
| D | Ireland | 0.4929*** (0.1368) | -0.0455* (0.0261) | 0.5540** (0.2225) | -0.0276 (0.0556) | 0.0007** (0.0004) | 0.5154*** (0.1150) | -0.0479* (0.0246) | 0.6461** (0.2170) | -0.0642 (0.0638) | 0.0008 (0.0006) | 145 |
| D | Italy | 0.5896*** (0.0496) | 0.0283* (0.0145) | 0.7061*** (0.0599) | -0.0476** (0.0220) | 0.0004** (0.0002) | 0.9769*** (0.0606) | -0.0069 (0.0256) | 0.9924*** (0.0913) | -0.0657* (0.0369) | 0.0004 (0.0003) | 145 |
| D | Japan | 0.2183** (0.1053) | -0.2154* (0.1274) | 0.3865** (0.1671) | -0.5047** (0.2302) | 0.0009** (0.0004) | 0.5259*** (0.1433) | -0.2260** (0.1161) | 1.0723*** (0.4230) | -0.4751** (0.2375) | 0.0006* (0.0003) | 145 |
| D | Netherlands | 0.5081*** (0.1009) | -0.0349* (0.0206) | 0.5503*** (0.1319) | -0.0533 (0.0382) | 0.0010*** (0.0002) | 0.5107*** (0.1738) | -0.0253* (0.0149) | 0.6894*** (0.2071) | -0.0033 (0.0482) | 0.0011*** (0.0004) | 145 |
| D | Norway | 0.4815*** (0.0802) | -0.0993* (0.0570) | 0.5603*** (0.1569) | -0.4354* (0.2500) | 0.0022*** (0.0004) | 0.5182*** (0.0733) | -0.0264* (0.0155) | 0.7156*** (0.1940) | -0.3042* (0.1782) | 0.0010** (0.0005) | 145 |
| D | Portugal | 0.5397*** (0.0723) | -0.0549* (0.0324) | 0.6637*** (0.1525) | -0.0748 (0.0667) | 0.0026*** (0.0028) | 0.5512*** (0.0757) | -0.0388* (0.0128) | 0.6513*** (0.1967) | -0.0377 (0.0647) | 0.0020*** (0.0008) | 144 |
| D | Singapore | 0.5658*** (0.1048) | 0.0212 (0.0252) | 0.2786 (0.1860) | -0.0504 (0.0540) | 0.0010*** (0.0003) | 0.6629*** (0.1117) | -0.0033 (0.0249) | 0.1974 (0.2682) | -0.0194 (0.0434) | 0.0006 (0.0004) | 145 |
| D | South Korea | 0.4615*** (0.0597) | -0.0689** (0.0314) | 0.5193*** (0.0886) | 0.5234 (0.6053) | 0.0017*** (0.0004) | 0.5371*** (0.1305) | -0.1028** (0.0529) | 0.9613*** (0.2141) | 0.1518 (0.381) | 0.0025*** (0.0007) | 145 |
| D | Spain | 0.5410*** (0.1072) | -0.0345* (0.0193) | 0.6498*** (0.1734) | -0.0125 (0.0419) | 0.0021*** (0.0007) | 0.6508*** (0.1097) | -0.0434* (0.0257) | 0.7359*** (0.1794) | -0.0096 (0.0493) | 0.0023*** (0.0007) | 145 |
| D | Sweden | 0.5611*** (0.0776) | -0.0563** (0.0275) | 0.6200*** (0.1299) | -0.0179 (0.1120) | 0.0010*** (0.0003) | 0.6756*** (0.1138) | -0.0601** (0.0291) | 0.9789*** (0.2319) | -0.0515 (0.1242) | 0.0006 (0.0004) | 145 |
| D | Switzerland | 0.6162*** (0.0639) | -0.0753* (0.0447) | 0.5732*** (0.1268) | -0.0472 (0.2333) | 0.0006*** (0.0002) | 0.5943*** (0.0913) | -0.1453** (0.0698) | 0.7924*** (0.1576) | 0.0151 (0.2776) | 0.0005 (0.0003) | 145 |
| D | United Kingdom | 0.4635*** (0.0490) | -0.0470** (0.0235) | 0.5018*** (0.1514) | -0.1163 (0.1295) | 0.0004 (0.0003) | 0.6308*** (0.0680) | -0.1885** (0.0822) | 0.7848*** (0.1744) | -0.0204 (0.1275) | 0.0016*** (0.0005) | 145 |
| D | United States | 0.4019*** (0.1174) | -0.0921*** (0.0291) | 0.4078*** (0.1253) | -0.0770 (0.0637) | 0.0018*** (0.0004) | 0.7431*** (0.1378) | -0.0746** (0.0297) | 0.8020*** (0.1989) | -0.0952 (0.0731) | 0.0007 (0.0005) | 145 |

Notes: Numbers in the table are coefficient estimates with HAC standard errors in parentheses. The *, **, and *** denote significance at 10%, 5%, and 1% levels respectively. The market classification is in S&P standards, and the superscripts "N" and "R" indicate estimations for normal (tranquil) and recessionary periods respective.

**Table 1.** *continued*

| Market Class. | Countries | Backward-Looking PC Model | | | | | Forward-Looking PC Model | | | | | Obs |
|---|---|---|---|---|---|---|---|---|---|---|---|---|
| | | $^N\pi_{t-1}$ | $^N U_{GAP}$ | $^R\pi_{t-1}$ | $^R U_{GAP}$ | C | $^N E_{t(t+1)}$ | $^N U_{GAP}$ | $^R E_{t(t+1)}$ | $^R U_{GAP}$ | C | |
| E | Brazil | 0.6328*** (0.1395) | 0.2578 (0.3374) | 0.7273*** (0.2099) | -0.2465 (0.6209) | 0.0052*** (0.0019) | 0.4545*** (0.0938) | 0.6443 (0.5081) | 0.5691*** (0.1420) | 0.0413 (1.1816) | 0.0130** (0.0052) | 106 |
| E | Chile | 0.7207*** (0.1395) | -0.0889** (0.0391) | 0.7888*** (0.1630) | -0.0273 (0.1157) | 0.0020*** (0.0006) | 0.9057*** (0.0484) | -0.0396 (0.0280) | 0.9235*** (0.1086) | -0.0797 (0.1388) | 0.0005 (0.0005) | 121 |
| E | China | 0.6191*** (0.0624) | -0.0248* (0.0144) | 0.6318*** (0.1178) | -0.0110 (0.0607) | 0.0011*** (0.0003) | 0.5816*** (0.1094) | -0.0452* (0.0260) | 0.7683*** (0.1497) | 0.0189 (0.0958) | 0.0004 (0.0005) | 145 |
| E | Czech | 0.5375*** (0.2952) | 0.0503 (0.1047) | 0.6318*** (0.3166) | -0.2434 (0.2155) | 0.0015** (0.0007) | 0.5118*** (0.2952) | -0.0404 (0.0839) | 0.7479** (0.3762) | -0.1972 (0.1752) | -0.0004 (0.0009) | 93 |
| E | Greece | 0.9730*** (0.0301) | 0.0002 (0.0259) | 0.8208*** (0.0721) | -0.0186 (0.0424) | 0.0008*** (0.0003) | 0.9324*** (0.1401) | -0.0104 (0.0526) | 0.2145 (0.1760) | -0.0226 (0.1198) | 0.0036*** (0.0013) | 145 |
| E | Hungary | 0.4646* (0.2752) | -0.2417 (0.2759) | 0.6549** (0.3236) | 0.0656 (0.4856) | 0.0052* (0.0028) | 0.5314*** (0.1082) | -0.4268 (0.3061) | 0.6037*** (0.1207) | 0.1812 (0.4789) | 0.0018 (0.0019) | 101 |
| E | India | 0.5483*** (0.1116) | -0.0390 (0.0249) | 0.5807*** (0.1350) | -0.0797 (0.0488) | 0.0055*** (0.0008) | 0.4984*** (0.1071) | -0.0896** (0.0444) | 0.6857*** (0.1477) | -0.0136 (0.1004) | 0.0068*** (0.0008) | 145 |
| E | Indonesia | 0.5892*** (0.0474) | -0.0284 (0.0732) | 0.6152*** (0.0932) | 0.1966 (0.7892) | 0.0060*** (0.0007) | 0.5592*** (0.0551) | -0.0013 (0.0631) | 0.6401*** (0.1434) | -0.0192 (0.9998) | 0.0063*** (0.0009) | 145 |
| E | Malaysia | 0.5217*** (0.1505) | -0.0391** (0.0182) | -0.0311 (0.0490) | -0.0479 (0.0815) | 0.0011* (0.0006) | 0.5452*** (0.1602) | -0.0626** (0.0263) | 0.0743 (0.3487) | -0.0495 (0.0835) | 0.0007 (0.0005) | 125 |
| E | Mexico | 0.6102*** (0.0894) | 0.2117 (0.2023) | 0.7227*** (0.1217) | 0.7032 (0.5722) | 0.0064*** (0.0021) | 0.7361*** (0.0852) | 0.1043 (0.1543) | 0.9072*** (0.1192) | 0.3224 (0.6259) | 0.0043** (0.0018) | 145 |
| E | Philippines | 0.6119*** (0.0392) | -0.0570** (0.0257) | 0.6645*** (0.2410) | -0.2853** (0.1173) | 0.0030*** (0.0005) | 0.4902*** (0.0750) | -0.0516** (0.0256) | 0.7646*** (0.1833) | -0.2211** (0.1025) | 0.0043*** (0.0008) | 145 |
| E | Poland | 0.6309** (0.2698) | -0.2459 (0.2949) | 0.7186** (0.3555) | 0.1151 (0.4710) | 0.0065* (0.0035) | 0.4408*** (0.0967) | -0.1187 (0.1273) | 0.6725*** (0.1188) | -1.5115 (1.3253) | 0.0023 (0.0028) | 109 |
| E | Russia | 0.5994** (0.2596) | 0.2381 (0.4535) | 0.6143** (0.2697) | -2.2745 (2.8064) | 0.0246* (0.0145) | 0.4576** (0.2128) | 0.1887 (0.4116) | 0.6822*** (0.3338) | -2.7840 (3.5023) | 0.0225** (0.0106) | 100 |
| E | SouthAftica | 0.5832*** (0.0549) | -0.0270 (0.0254) | 0.6998*** (0.0869) | 0.0934 (0.0625) | 0.0027*** (0.0005) | 0.5118*** (0.0942) | -0.0350 (0.0251) | 0.9690*** (0.1094) | 0.0988* (0.0538) | 0.0017* (0.0009) | 145 |
| E | Taiwan | 0.5311** (0.2633) | -0.2679*** (0.0649) | 0.5895** (0.2446) | -0.2745 (0.1709) | 0.0020*** (0.0004) | 0.5265** (0.1946) | -0.2569*** (0.0858) | 0.6322*** (0.2383) | -0.1971 (0.2506) | 0.0015*** (0.0004) | 145 |
| E | Thailand | 0.4898*** (0.1302) | -0.0464 (0.0358) | 0.5730*** (0.1682) | -0.0638 (0.0809) | 0.0017*** (0.0004) | 0.5716*** (0.1475) | -0.0837 (0.0525) | 0.7290*** (0.1950) | -0.0924 (0.0938) | 0.0014** (0.0004) | 144 |
| F | Turkey | 0.6434*** (0.0915) | 0.0346 (0.1650) | 0.7029*** (0.1482) | -0.0884 (0.3256) | 0.0077*** (0.0024) | 0.6954*** (0.0368) | -0.0712 (0.0882) | 0.8248*** (0.0571) | -0.0219 (0.2039) | 0.0056*** (0.0018) | 145 |
| F | Argentina | 0.6617*** (0.1031) | -0.3075 (0.2209) | 1.0092*** (0.2871) | 0.4933 (1.2137) | 0.0187** (0.0088) | 0.6110*** (0.0461) | 0.0104 (0.1737) | 0.6912*** (0.0590) | 0.0944 (0.2598) | -0.0011 (0.0015) | 129 |
| F | Romania | 0.1873 (0.2207) | 0.0304 (0.2335) | 0.4118 (0.2714) | -1.0172 (0.9186) | 0.0250*** (0.0107) | 0.4559*** (0.1008) | 0.0402 (0.1106) | 0.5862** (0.2366) | 0.8197 (0.6136) | 0.0027 (0.0039) | 102 |
| F | Venezuela | 0.6938*** (0.1100) | 0.0253 (0.0541) | 0.8090*** (0.1287) | -0.3190* (0.1713) | 0.0109*** (0.0029) | 0.4020*** (0.1053) | 0.0740 (0.0819) | 0.5337*** (0.1249) | 0.0695 (0.2545) | 0.0221*** (0.0047) | 112 |

Notes: Numbers in the table are coefficient estimates with HAC standard errors in parentheses. The *, **, and *** denote significance at 10%, 5%, and 1% levels respectively. The market classification is in S&P standards, and the superscripts "N" and "R" indicate estimations for normal (tranquil) and recessionary periods respective.

Table 1 derives several plausible results. In the left hand side of the table, estimates of the backward-looking model show that the Phillips relation ($^N U_{GAP}$) works in the majority of developed countries during normal (tranquil) economic periods, but its significance remains limited at 10% level. Apparently, the coefficient gains statistically more significance in Canada, South Korea, Sweden, United Kingdom, and the United States in individual cases, while it completely fails in a few developed countries, such as Belgium, Denmark, Finland, Italy, and Singapore. It remains unclear whether the Phillips trade-off derives positive significant results in Italy. Del Boca et al (2010) also underlined the failure of the Phillips relationship in their study. To generalize, when the outliers are excluded, the average backward-looking Phillips coefficient ($^N U_{GAP}$) appears around -0.07 for the developed market sample during normal (tranquil) periods.

The backward-looking NCPC, even the forward-looking NCPC in the right-hand side of the Table 1, fails to work in the majority of emerging and frontier markets during both tranquil and recessionary periods. This supports the findings Paul (2009) and Sovbetov and Kaplan (2019), who observe that the rela-tionship is often evasive or absent in less-developed and crisis-prone markets due to a lack of smoothly operating macroeconomic foundations and the tranquillity of the economic environment. It is worth noting that the majority of the sample of emerging markets is comprised of Latin American and Asian countries that have experienced many sovereign debt crises and currency crashes during 1980-1999. [5] In addition, NCPC also fails to work in Greece, Romania, and Turkey due to undulant economic conditions. These two countries have experienced about 35-40 quarters of recessions just during 1980-1990 (Appendix-A).

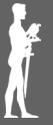

Moreover, the backward-looking fraction of inflation ($^{N}\pi_{t-1}$) appears statistically significant at 1% level in the majority of the sample during normal economic periods, with exception of Romania, where lagged inflation ambiguously fails to be significant. Its average magnitude in developed, emerging, and frontier markets is about 50.73%, 60.63%, and 67.78%, respectively. This, once again, shows that the most developed countries are less backward-looking compared to emerging and frontier ones.

On the other hand, the table plainly shows that the Phillips relation ($^{R}U_{GAP}$) collapses, and the backward-looking fraction of inflation ($^{R}\pi_{t-1}$) remarkably increases in magnitude within whole sample countries without any loss in significance during recessionary periods. The average coefficient of past inflation scales up from 50.73% to 56.34% in developed markets, from 60.63% to 67.10% in emerging markets, and from 67.78% to 90.91% in frontier markets. This indicates that markets become more inflation-sensitive during recessionary periods, as the backward-looking coefficient gains weight and significance.

On the right hand side of the table, results of forward-looking model show that the Phillips relation ($^{N}U_{GAP}$) works in the majority of developed countries during normal economic periods, with better significance levels compared to backward-looking cases. It is clear that the Phillips coefficient ($^{N}U_{GAP}$) gains remarkable significance especially in Austria, Belgium, Denmark, Finland, France, Germany, Japan, Switzerland, and United States. When insignificant results are excluded, the average Phillips coefficient ($^{N}U_{GAP}$) appears the same as it was in the backward-looking case, -0.07, for the developed market sample during normal (tranquil) periods. In the cases of emerging and frontier countries, forward-looking model generates alike results as backward-looking one. The forward-looking NCPC seems not to work in these samples.

Moreover, the forward-looking fraction of inflation [$^{N}Et_{(\pi t+1)}$] appears statistically significant at 1% level in the majority of the samples during normal economic periods, without any exceptions. Its average magnitude in developed, emerging, and frontier markets is about 62.05%, 58.53%, and 48.96% respectively. This indicates that developed countries are more forward-looking than emerging and frontier ones. Besides, Phillip relation ($^{R}U_{GAP}$) fails to be valid throughout the whole sample during recessionary periods, and a forward-looking fraction of inflation [$^{N}Et_{(\pi t+1)}$] considerably increases in magnitude within all sample countries without any loss in significance, even in emerging and frontier markets. The average coefficient of expected inflation rises from 62.05% to 80.09% in developed markets; from 58.53% to 74.13% in emerging markets; and from 48.96% to 60.37% in frontier markets. This indicates that the dominance of the expected inflation in the forward-looking model increases in recessionary periods comparing to tranquil (normal) periods in all the sampled countries, regardless of their market classification. In other words, countries become more inflation-sensitive during recessionary periods, as forward-looking coefficient gains weight and significance.

Notice that both backward- and forward-looking models estimates overall increase in weight and significance of inflation factor (past inflation in backward-looking case and expected future inflation in forward-looking case) during recessionary periods. Apparently, this picks up due to two issues. First, the Phillips coefficient loses its significance during recession, thus, current inflation becomes more sensitive to past or expected future inflations.

---

Second, neither models incorporate past inflation and expected future inflation variables simultaneously in a hybrid form. Thus, the study cannot clearly conclude whether markets become more backward- or forward-looking during recessionary periods. The findings only show that inflation becomes more sensitive to its past or expected future values during recessionary periods and the Phillips relation demises.

In addition, the sample of emerging and frontier markets are predominantly comprised of inflation-prone fragile countries that have experienced many non-growth periods since the beginning of the analysis period (1980). Developed markets, however, have relatively fewer recessionary periods. This also might have some impact on limited increases in past inflation coefficients of backward-looking model in developed markets during recessionary periods, while the coefficient increases remarkably in emerging and frontier markets.

## CONCLUSION

This study examines the behaviour of NCPC during tranquil and recessionary periods and documents several findings. Based on the results of this research, first of all, the study finds that both backward- and forward-looking NCPC models work in the majority of developed markets during tranquil periods. However, the significance of the backward-looking model is much weaker compared to the forward-looking model.

Second, both backward- and forward-looking NCPC models fail to work in the majority of emerging and frontier markets, even in tranquil periods. This is because they are predominantly comprised of inflation-prone fragile countries that have experienced many recessionary periods since the beginning of the analysis period. This supports the findings Paul (2009) and Sovbetov and Kaplan (2019), who conclude that the relationship is often evasive or absent in less developed and crisis-prone countries due to a lack of well-established and smoothly operating macroeconomic foundations.

Third, both backward- and forward-looking NCPC models completely collapse, deriving statistically insignificant Phillips coefficient during recessionary periods in the whole sample. This shows that the tranquillity of economic environment significantly matters for the Phillips trade-off to work smoothly.

Fourth, the study documents that developed countries tend to be more forward-looking (less backward-looking) comparing to emerging and frontier ones during tranquil periods.

Fifth, during recessionary periods both backward- and forward-looking fractions of inflation remarkably increase in magnitude within whole sample countries without any loss in significance. This indicates that markets become more inflation-sensitive during recessionary periods. Apparently, this picks up for two reasons. First, the Phillips coefficient loses its significance during recessions, thus, current inflation becomes more sensitive to past or expected future inflation. Second, neither models incorporate past inflation and expected future inflation variables simultaneously in a hybrid form. Thus, the study cannot clearly conclude whether markets become more backward- or forward-looking during recessionary periods or not. This should be considered by future researches in related fields.

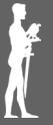

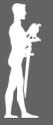

## APPENDIX

**Table A1.** *Country Codes and Number of Recessions Different Time Periods*

| Country Name | Code | 1980-1990 40 quarters | 1990-2000 40 quarters | 1980-2016 145 quarters | 1990-2016 105 quarters | 2000-2016 65 quarters |
|---|---|---|---|---|---|---|
| Argentina | AG | 22 | 17 | 55 | 33 | 16 |
| Australia | AU | 9 | 3 | 15 | 6 | 3 |
| Germany | BD | 13 | 13 | 42 | 29 | 16 |
| Belgium | BG | 6 | 6 | 23 | 17 | 11 |
| Brazil | BR | 19 | 16 | 51 | 32 | 16 |
| Canada | CH | 8 | 3 | 11 | 3 | 0 |
| Chile | CL | 10 | 8 | 32 | 22 | 14 |
| China | CN | 11 | 4 | 23 | 12 | 8 |
| Czech Republic | CZ | - | 13 | 24 | 24 | 11 |
| Denmark | DK | 15 | 11 | 50 | 35 | 24 |
| Spain | ES | 9 | 6 | 32 | 23 | 17 |
| Finland | FN | 5 | 15 | 43 | 38 | 23 |
| France | FR | 2 | 5 | 22 | 20 | 15 |
| Greece | GR | 21 | 14 | 75 | 54 | 40 |
| Hungary | HN | - | 17 | 27 | 27 | 10 |
| Indonesia | ID | 9 | 7 | 17 | 8 | 1 |
| India | IN | 10 | 9 | 24 | 14 | 5 |
| Ireland | IR | 15 | 11 | 47 | 32 | 21 |
| Italy | IT | 7 | 13 | 47 | 40 | 27 |
| Japan | JP | 6 | 16 | 46 | 40 | 24 |
| South Korea | KO | 4 | 4 | 11 | 7 | 3 |
| Mexico | MX | 16 | 4 | 31 | 15 | 11 |
| Malaysia | MY | 3 | 3 | 13 | 10 | 7 |
| Netherlands | NL | 11 | 3 | 31 | 20 | 17 |
| Norway | NW | 12 | 12 | 44 | 32 | 20 |
| Austria | OE | 10 | 2 | 32 | 22 | 20 |
| Philippines | PH | 11 | 7 | 21 | 10 | 3 |
| Poland | PO | - | 7 | 15 | 15 | 8 |
| Portugal | PT | 4 | 7 | 37 | 33 | 26 |
| Romania | RM | 19 | 23 | 56 | 37 | 14 |
| Russia | RS | - | 26 | 37 | 37 | 11 |
| South Africa | SA | 11 | 12 | 28 | 17 | 5 |
| Sweden | SD | 9 | 11 | 32 | 23 | 12 |
| Singapore | SP | 3 | 5 | 24 | 21 | 16 |
| Switzerland | SW | 6 | 13 | 31 | 25 | 12 |
| Thailand | TH | 10 | 8 | 29 | 19 | 11 |
| Turkey | TK | 19 | 18 | 47 | 28 | 10 |
| Taiwan | TW | 6 | 4 | 30 | 24 | 20 |
| United Kingdom | UK | 5 | 6 | 18 | 13 | 7 |
| United States | US | 6 | 2 | 18 | 12 | 10 |
| Venezuela | VE | 20 | 13 | 53 | 33 | 20 |

**Notes:** Numbers in the table show the quarter numbers with negative GDP growth (recession). The "-" denote missing data.

**Table A2**. *Results of Unit Root Tests for Series of backward- and forward-looking EAPC*

|  | ADF (intercept) | | | PP (intercept) | | |
|---|---|---|---|---|---|---|
|  | CPI | EI | U_U' | CPI | EI | U_U' |
| AG | 0.0791 (L:2\|N:126) | 0.0508 (L:2\|N:126) | 0.0001 (L:0\|N:144) | 0.0000 (B:7\|N:128) | 0.0000 (B:7\|N:128) | 0.0000 (B:2\|N:144) |
| AU | 0.0008 (L:1\|N:143) | 0.0000 (L:0\|N:144) | 0.0010 (L:2\|N:142) | 0.0000 (B:8\|N:144) | 0.0000 (B:7\|N:144) | 0.0423 (B:6\|N:144) |
| BD | 0.1003 (L:3\|N:141) | 0.0355 (L:3\|N:141) | 0.0419 (L:4\|N:140) | 0.0000 (B:10\|N:144) | 0.0000 (B:10\|N:144) | 0.0008 (B:9\|N:144) |
| BG | 0.0000 (L:0\|N:144) | 0.0350 (L:1\|N:143) | 0.4779 (L:3\|N:141) | 0.0000 (B:9\|N:144) | 0.0001 (B:9\|N:144) | 0.0718 (B:3\|N:144) |
| BR | 0.2191 (L:2\|N:103) | 0.115 (L:3\|N:103) | 0.0000 (L:8\|N:136) | 0.0993 (B:1\|N:105) | 0.2073 (B:6\|N:106) | 0.0000 (B:8\|N:144) |
| CH | 0.0045 (L:4\|N:140) | 0.0040 (L:4\|N:140) | 0.0000 (L:1\|N:143) | 0.0000 (B:10\|N:144) | 0.0000 (B:10\|N:144) | 0.0172 (B:5\|N:144) |
| CL | 0.7175 (L:7\|N:137) | 0.7916 (L:7\|N:137) | 0.0002 (L:1\|N:119) | 0.0000 (B:8\|N:144) | 0.0001 (B:7\|N:144) | 0.0007 (B:4\|N:120) |
| CN | 0.0173 (L:3\|N:141) | 0.0106 (L:2\|N:142) | 0.0431 (L:0\|N:144) | 0.0000 (B:8\|N:144) | 0.0084 (B:4\|N:144) | 0.0241 (B:4\|N:144) |
| CZ | 0.0677 (L:3\|N:96) | 0.0002 (L:0\|N:100) | 0.0081 (L:5\|N:87) | 0.0000 (B:3\|N:99) | 0.0000 (B:17\|N:100) | 0.0509 (B:3\|N:92) |
| DK | 0.0361 (L:4\|N:140) | 0.0273 (L:3\|N:141) | 0.5460 (L:1\|N:143) | 0.0000 (B:10\|N:144) | 0.0000 (B:9\|N:144) | 0.2476 (B:7\|N:144) |
| ES | 0.3683 (L:7\|N:137) | 0.0393 (L:7\|N:137) | 0.6306 (L:1\|N:143) | 0.0000 (B:8\|N:144) | 0.0003 (B:10\|N:144) | 0.3053 (B:7\|N:144) |
| FN | 0.0154 (L:4\|N:140) | 0.0120 (L:4\|N:140) | 0.0001 (L:4\|N:140) | 0.0000 (B:10\|N:144) | 0.0001 (B:10\|N:144) | 0.0282 (B:9\|N:144) |
| FR | 0.0308 (L:11\|N:133) | 0.0239 (L:0\|N:144) | 0.0159 (L:1\|N:143) | 0.0055 (B:9\|N:144) | 0.0298 (B:12\|N:144) | 0.0371 (B:4\|N:144) |
| GR | 0.3312 (L:4\|N:140) | 0.5571 (L:4\|N:140) | 0.0000 (L:8\|N:136) | 0.0000 (B:10\|N:144) | 0.0000 (B:11\|N:144) | 0.036 (B:7\|N:144) |
| HN | 0.3649 (L:3\|N:141) | 0.4055 (L:6\|N:138) | 0.0098 (L:1\|N:99) | 0.0000 (B:10\|N:144) | 0.0000 (B:11\|N:144) | 0.0714 (B:0\|N:100) |
| ID | 0.0000 (L:0\|N:144) | 0.0000 (L:1\|N:143) | 0.0000 (L:4\|N:140) | 0.0000 (B:1\|N:144) | 0.0018 (B:9\|N:144) | 0.0015 (B:2\|N:144) |
| IN | 0.0031 (L:3\|N:141) | 0.0009 (L:4\|N:140) | 0.0000 (L:4\|N:140) | 0.0000 (B:9\|N:144) | 0.0000 (B:10\|N:144) | 0.0000 (B:10\|N:144) |
| IR | 0.0067 (L:4\|N:140) | 0.0011 (L:4\|N:140) | 0.1113 (L:2\|N:142) | 0.0000 (B:7\|N:144) | 0.0000 (B:3\|N:144) | 0.0652 (B:8\|N:144) |
| IT | 0.0061 (L:8\|N:136) | 0.0001 (L:9\|N:135) | 0.0024 (L:0\|N:144) | 0.0673 (B:9\|N:144) | 0.0881 (B:10\|N:144) | 0.0027 (B:6\|N:144) |
| JP | 0.0001 (L:1\|N:143) | 0.0086 (L:2\|N:142) | 0.6762 (L:0\|N:144) | 0.0000 (B:9\|N:144) | 0.0000 (B:8\|N:144) | 0.3394 (B:6\|N:144) |
| KO | 0.0001 (L:3\|N:141) | 0.0149 (L:2\|N:142) | 0.1671 (L:2\|N:142) | 0.0000 (B:8\|N:144) | 0.0000 (B:9\|N:144) | 0.0440 (B:5\|N:144) |

**Notes:** The numbers in the table are rejection probabilities of the null hypotheses of ADF and PP tests including intercept. Probabilities below 0.10 denote rejection of these null hypotheses, thus, confirm stationarity of the CPI (inflation), EI (expected inflation), and U_U' (unemployment gap) series of related countries. The lag and observation parameters are presented in the parentheses where "L", "B", and "N" denote lag length, Newey-West bandwidth using Bartlett kernel, and observation number respectively. The lag length is determined by Schwarz Information Criterion (SIC) under maximum lag length specification of 13. Initials are given in table A1.

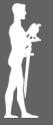

**Table A2**. *(continues)*

| | ADF (intercept) | | | PP (intercept) | | |
|---|---|---|---|---|---|---|
| | CPI | EI | U_U' | CPI | EI | U_U' |
| MX | 0.0415 (L:0\|N:144) | 0.3655 (L:9\|N:135) | 0.0414 (L:4\|N:140) | 0.0672 (B:7\|N:144) | 0.0901 (B:6\|N:144) | 0.0567 (B:6\|N:144) |
| MY | 0.0000 (L:0\|N:144) | 0.0027 (L:1\|N:143) | 0.0006 (L:0\|N:124) | 0.0000 (B:4\|N:144) | 0.0000 (B:8\|N:144) | 0.001 (B:4\|N:124) |
| NL | 0.0274 (L:3\|N:141) | 0.0040 (L:4\|N:140) | 0.1580 (L:12\|N:132) | 0.0000 (B:10\|N:144) | 0.0000 (B:10\|N:144) | 0.0557 (B:5\|N:144) |
| NW | 0.0602 (L:3\|N:141) | 0.0000 (L:3\|N:141) | 0.2180 (L:0\|N:144) | 0.0000 (B:9\|N:144) | 0.0000 (B:10\|N:144) | 0.0848 (B:5\|N:144) |
| OE | 0.0059 (L:4\|N:140) | 0.0101 (L:3\|N:141) | 0.0842 (L:0\|N:144) | 0.0000 (B:9\|N:144) | 0.0000 (B:9\|N:144) | 0.1008 (B:1\|N:144) |
| PH | 0.0002 (L:2\|N:142) | 0.0005 (L:2\|N:142) | 0.0001 (L:4\|N:140) | 0.0000 (B:7\|N:144) | 0.0000 (B:7\|N:144) | 0.0000 (B:9\|N:144) |
| PO | 0.0964 (L:9\|N:108) | 0.2312 (L:6\|N:112) | 0.0066 (L:2\|N:106) | 0.0001 (B:3\|N:117) | 0.0076 (B:3\|N:118) | 0.0768 (B:6\|N:108) |
| PT | 0.5028 (L:7\|N:137) | 0.3723 (L:7\|N:137) | 0.1767 (L:1\|N:143) | 0.0000 (B:10\|N:144) | 0.0005 (B:10\|N:144) | 0.0000 (B:9\|N:144) |
| RM | 0.0000 (L:0\|N:101) | 0.0000 (L:0\|N:102) | 0.0001 (L:4\|N:108) | 0.0000 (B:8\|N:101) | 0.0000 (B:8\|N:102) | 0.0022 (B:7\|N:112) |
| RS | 0.1050 (L:2\|N:97) | 0.1158 (L:1\|N:99) | 0.0004 (L:4\|N:100) | 0.0002 (B:3\|N:99) | 0.0410 (B:2\|N:100) | 0.0006 (B:7\|N:104) |
| SA | 0.0364 (L:2\|N:142) | 0.2099 (L:2\|N:142) | 0.0000 (L:5\|N:139) | 0.0000 (B:7\|N:144) | 0.0007 (B:9\|N:144) | 0.0000 (B:10\|N:144) |
| SD | 0.1198 (L:3\|N:141) | 0.0463 (L:3\|N:141) | 0.1913 (L:1\|N:143) | 0.0000 (B:9\|N:144) | 0.0002 (B:9\|N:144) | 0.1991 (B:7\|N:144) |
| SP | 0.0000 (L:0\|N:144) | 0.0035 (L:3\|N:141) | 0.0000 (L:1\|N:143) | 0.0000 (B:4\|N:144) | 0.0000 (B:3\|N:144) | 0.0159 (B:10\|N:144) |
| SW | 0.0496 (L:4\|N:140) | 0.1493 (L:3\|N:141) | 0.0000 (L:1\|N:143) | 0.0000 (B:10\|N:144) | 0.0000 (B:9\|N:144) | 0.0228 (B:7\|N:144) |
| TH | 0.0000 (L:0\|N:144) | 0.0003 (L:1\|N:143) | 0.0000 (L:4\|N:140) | 0.0000 (B:7\|N:144) | 0.0000 (B:6\|N:144) | 0.0002 (B:1\|N:144) |
| TK | 0.5735 (L:3\|N:141) | 0.6090 (L:2\|N:142) | 0.0000 (L:8\|N:136) | 0.0000 (B:9\|N:144) | 0.0099 (B:7\|N:144) | 0.0000 (B:10\|N:144) |
| TW | 0.0000 (L:3\|N:141) | 0.0042 (L:6\|N:138) | 0.1369 (L:5\|N:139) | 0.0000 (B:7\|N:144) | 0.0000 (B:6\|N:144) | 0.1121 (B:9\|N:144) |
| UK | 0.0117 (L:4\|N:140) | 0.0089 (L:4\|N:140) | 0.0413 (L:2\|N:142) | 0.0000 (B:10\|N:144) | 0.0000 (B:9\|N:144) | 0.0064 (B:8\|N:144) |
| US | 0.0002 (L:2\|N:142) | 0.0000 (L:4\|N:140) | 0.0295 (L:5\|N:139) | 0.0000 (B:7\|N:144) | 0.0000 (B:3\|N:144) | 0.0022 (B:6\|N:144) |
| VE | 0.0074 (L:0\|N:143) | 0.4144 (L:0\|N:111) | 0.0000 (L:4\|N:140) | 0.0095 (B:8\|N:143) | 0.2463 (B:2\|N:111) | 0.0000 (B:9\|N:144) |

# UZROCI NEUSPEHA FILIPSOVE KRIVE: DA LI JE BITNO DA EKONOMSKO OKRUŽENJE NE OSCILIRA?

**Rezime:**

Iako je empirijska literatura koja se odnosi na Filipsovu krivu (the Phillips curve) značajnog obima, i dalje ne postoji konsenzus o validnosti i stabilnosti iste. U literaturi se navodi da je Filipsov odnos nestalan i da je drugačiji od zemlje do zemlje i u različitim vremenskim periodima; statistički odnos koji se u toku jedne decenije (u nekoj zemlji) čini jakim, može biti slab u narednoj/nekoj drugoj. Razlozi za ovu nestalnost mogu biti osnova za osobenosti neke zemlje i njenog ekonomskog okruženja. Kako bismo se pozabavili ovom temom, u radu smo detaljno istražili Filipsov odnos u 41 zemlji tokom perioda 1980-2016, obraćajući pažnju na dinamiku inflacije tokom perioda bez većih oscilacija i recesije. Kao rezultat, u radu je zaključeno da Filipsov odnos varira, u zavisnosti od zemlje i vremenskog perioda. Dokazano je da je taj odnos važeći za većinu razvijenih zemalja, dok nije primenjiv u zemljama u razvoju i nerazvijenim zemljama, tokom perioda bez većih oscilacija. S druge strane, odnos je nepostojeći tokom perioda recesije, čak i na razvijenim tržištima. Ovo dokazuje da je period bez većih oscilacija u ekonomskom okruženju od izuzetnog značaja, kako bi Filipsov balans funkcionisao bez problema. Štaviše, frakcije – očekivane inflacije i inflacije u prethodnom periodu, značajno se uvećavaju tokom perioda recesije kao rezultat toga što Filipsov koeficijent gubi na značaju u okviru modela. Ovo ukazuje na činjenicu da su tržišta osetljivija na inflaciju tokom ovih perioda.

**Ključne reči:**

Filipsova kriva, inflacija, recesija, stabilnost